\documentclass[12pt,preprint]{aastex}

\shorttitle{Shock Acceleration} \shortauthors{Liu et al.}

\begin{document}

\title{Stochastic Electron Acceleration in Shell-Type Supernova Remnants}
\author{Siming Liu\altaffilmark{1}, Zhong-Hui Fan\altaffilmark{2,3},
Christopher L. Fryer\altaffilmark{1, 4}, Jian-Min
Wang\altaffilmark{2,5}, and Hui Li\altaffilmark{1}}

\altaffiltext{1}{Los Alamos National Laboratory, Los Alamos, NM
87545; liusm@lanl.gov} \altaffiltext{2}{Institute of
High Energy Physics, Chinese Academy of Sciences, Beijing 100049,
China}
\altaffiltext{3}{Department of Physics, Yunnan University, Kunming
650091, Yunnan, China} \altaffiltext{4}{Physics Department, The
University of Arizona, Tucson, AZ 85721}
\altaffiltext{5}{Theoretical Physics Center for Science Facilities,
Chinese Academy of Sciences}


\begin{abstract}
We study the stochastic electron acceleration by fast mode waves in
the turbulent downstream of weakly magnetized collisionless
astrophysical shocks. The acceleration is most efficient in a
dissipative layer, and the model characteristics are determined by
the shock speed,  density, magnetic field, and turbulence decay
length. The model explains observations of shell-type supernova
remnants RX J1713.7-3946 and J0852.0-4622 and can be tested by
observations in hard X-rays with the {\it HXMT} and {\it NuSTAR} or
$\gamma$-rays with the {\it GLAST}.
\end{abstract}


\section{Introduction}
Recent observations of high-energy emission from a few shell-type
supernova remnants (STSNRs) advance our understanding of the
underlying physical processes significantly \citep{e02}. These
observations not only give high quality emission morphology and spectra
 \citep{u03, a04, c04, h05, t08a}, which lead to tight
upper limits on thermal X-rays, but also discover X-ray variability
on a timescale of a few months \citep{u07}, which is intimately
connected to processes near the shock front (SF). The radio to X-ray
emissions are  produced by relativistic electrons through the
synchrotron processes. The nature of the TeV emission is still a
matter of debate. The major challenges to the hadronic scenario
include: 1) the lack of correlation between the TeV
emission and molecular cloud distribution; 2) the requirement of
very high mean target gas density, supernova explosion energy,
and/or proton acceleration efficiency, which implies efficient
magnetic field amplification in the context of the diffusive shock
acceleration and strong suppression of the electron acceleration \citep{lb00, bv06, b08}; 3) the good correlation between the
X-ray and TeV emission \citep{p08, a06, a07a}, which
suggests a leptonic scenario. Because secondary leptons from
hadronic processes won't produce as much radiation power as
$\gamma$-rays from neutral pion ($\pi^0$) decays \citep{a07b}, the
fact that the X-ray luminosity is slightly higher than the TeV
luminosity requires that most leptons be accelerated from
the background plasma.

Compared to the hadronic scenario, the leptonic scenario has much
fewer parameters and is well-constrained by the observed radio,
X-ray, and TeV spectra. In the simplest one-zone model, the TeV
emission is produced through the inverse Compton (IC) scattering of
the Galactic interstellar radiation by the same relativistic
electrons producing the radio and X-rays \citep{p06}. The TeV
spectrum can be used to infer the high-energy electron distribution.
One then needs to fit both the synchrotron spectral shape and flux
level by adjusting the magnetic field alone. TeV observations always
see spectral softening with the increase of the photon energy,
suggesting a spectral cutoff \citep{a07a, a07b}. There are also
indications that the synchrotron spectrum cuts off in the hard X-ray
band \citep{t08a, t08b, u07}.
%Therefore the magnetic field and high energy electron distribution
%can be determined by the X-ray and TeV spectra.
A single power-law model with an exponential cutoff for the electron
distribution has difficulties in reproducing the broad TeV spectrum.
Models with more gradual cutoffs lead to much better fits.
Relativistic particle distributions with gradual high-energy cutoffs
are a natural consequence of the stochastic acceleration
(SA) by plasma waves in magnetized turbulence \citep{pp95, bld06,
sp08}. The particle distribution is determined by the interplay
among the acceleration, cooling, and escape processes with the
cutoff shape determined by the energy dependence of the ratio of the
acceleration to cooling or escape rate.
%As mentioned above, cooling is unimportant in determining the TeV emitting electron distribution
Given the low magnetic fields derived from leptonic models, the
radiative cooling time is much longer than the remnant lifetime.
%Such a spectral cutoff has to be generic to the particle acceleration processes.
The gradual high-energy cutoff  has to be caused by  the balance
between the acceleration and escape processes.

We consider the evolution of weakly magnetized turbulence in the
downstream of strong non-relativistic shocks and study the SA by fast mode waves. The large
scale turbulence cascades following the Kolmogorov
phenomenology \citep{b02, j08} until it reaches the scale, where the
eddy turnover time becomes comparable to the period of fast mode
waves. The Iroshnikov-Kraichnan phenomenology prevails on even
smaller scales. The collisionless damping of plasma waves sets in at
the coherent length of the magnetic field, where the eddy turnover
time is comparable to the period of Alfv\'{e}n waves. Fast mode
waves propagating nearly parallel to the mean magnetic field can
survive the transit-time damping (TTD) by the thermal background
particles, and accelerate some electrons through cyclotron
resonances to a power-law high-energy distribution,
which cuts off at the energy, where the particle gyro-radius reaches
the coherent length of the magnetic field \citep{l06}. Acceleration
of higher energy electrons by the nearly isotropic fast mode waves
on larger scales leads to a gradual cutoff. SA of
electrons can naturally account for  emissions from the STSNRs with
a weak magnetic field. Future observations in the hard X-ray and MeV
bands may test the model. The particle acceleration model is
described in \S\ \ref{model} and applied to SNRs RX J1713-3946 and J0852.0-4622 to derive quantitative results in \S\ \ref{app}. \S\
\ref{con} gives the conclusions.

\section{Turbulence Cascade, Wave Damping, and Stochastic Particle Acceleration}
\label{model}

The collisionless nature of weakly magnetized shocks implies
difficulties in converting the free energy in the system into heat
and energetic particles instantaneously at a narrow SF. Instead, the
high magnetic Reynolds number implies strong turbulence.
Previous studies of diffusive shocks focus on its roles in making high-energy particles
crossing the SF repeatedly. The turbulence can also carry a
significant fraction of the released free energy and decay
gradually. The turbulence generation mechanism is not well understood. It is usually attributed to the Weibel instability \citep{W59} of particle streaming and may also depend on the properties of the upstream plasma. We here assume that the turbulence is isotropic and generated at a scale $L$ with an eddy speed $u$. The free energy dissipation rate is then given by $Q=
C_1\rho u^3/L\,, $ where $C_1\sim1$ is dimensionless.
For strong non-relativistic shocks with the shock frame upstream speed $U$ much higher than the speed
of the upstream fast mode waves $v_F = (v_A^2+5v_S^2/3)^{1/2}$,
mass, momentum, and energy conservations across the SF require $U^2
= 5 v_S^2 + 5 u^2+ 2v_A^2+U^2/16\,,$ where the Alfv\'{e}n speed
$v_A=(B^2/4\pi\rho)^{1/2}$, the isothermal sound speed
$v_S = (P/\rho)^{1/2}$, and $B$, $\rho$, $P$ are the magnetic
field, mass density, pressure, respectively.
 The shock structure can be complicated due to the presence of strong turbulence, and the speeds $v_S$, $v_A$, and $u$ should be considered as averaged quantities over $\sim L$, which also corresponds to the energy dissipation scale.
%may not be well defined at a given distance to the SF. Since the free energy dissipation is most
%efficient near the SF, we here focus on a simple one zone model with $u\ge v_F$ and a turbulence decay length $L$. Thus the corresponding quantities should be interpreted as `averaged' values in this dissipation range.
The excitation of plasma waves should be very efficient in the supersonic phase with $u\ge v_F$, and the waves may prevail in the downstream accelerating some particles in the background plasma to very high energies.
%and particles may be accelerated to very high energies through resonance interactions with these waves. Away from the SF, the downstream flow becomes subsonic with $u<v_F$ and the SA becomes much inefficient due to difficulties in exciting high speed waves.

The eddy turnover speed and time are given respectively by
$v^2_{edd}(k)=4\pi W(k)k^3$ and $\tau_{edd}(k) = 2\pi /C_1kv_{edd}
%=\pi^{1/2}(C_1^2W k^5)^{-1/2}
$, where $W(k) =
%(u^2/4\pi)(2\pi/L)^{2/3} k^{-11/3} =
(4\pi)^{-1} (2\pi
Q/C_1\rho)^{2/3} k^{-11/3}$, $k=2\pi/l$, and $l$ are the isotropic turbulence power
spectrum,  wave number, and eddy size, respectively. At the turbulence generation scale $k_m = 2\pi/L$, $v_{edd}=u$,
%$Q=2 C_1\rho [4\pi W^3k^{11}]^{1/2}= \rho v^2_{edd}(k)/ \tau_{edd}(k)$,
and the total
turbulence energy is given by $\int W(k) 4\pi k^2 {\rm d} k = (3/2)
u^2$. The MHD wave period is given by $\tau_F(k)=2\pi /v_Fk$. Then
the transition scale from the Kolmogorov to Kraichnan phenomenology $k_t$
occurs at $\tau_F(k_t) = \tau_{edd}(k_t)$, % or $v_F = C_1v_{edd}(k_t)$,
which gives $k_t = (C_1u/v_F)^3k_m\,.$ For
$k>k_t$, the turbulence spectrum in the inertial range is given by
\begin{equation}
W(k) %= (u^2/4\pi)k_m^{2/3} k_t^{-11/3} (k/k_t)^{-7/2}
%= (u^2/4\pi)k_m^{2/3} k_t^{-1/6}k^{-7/2}
=(4\pi)^{-1}
(v_F/C_1)^{1/2}u^{3/2}k_m^{1/2}k^{-7/2}\,. \label{IK}
\end{equation}
The collisionless damping starts at the coherent length of the
magnetic field $l_d = 2\pi/k_d$, where the period of Alfv\'{e}n
waves $2\pi/kv_A$ is comparable to the eddy turnover time,
%$v_A^2 = C_1^2 4\pi W k_d^3
%= v_F^{1/2} (C_1u)^{3/2}k_m^{1/2} k_d^{-1/2} $, which implies
$k_d = (C_1^3u^{3}v_F^{}/v_A^4) k_m\, .$

Both $v_A$ and $v_S$ may be lower than $u$ near the SF. If
$u\gg v_A\gg v_S$ at the SF, fast, slow, and Alfv\'{e}n waves can all be
excited, and the consequent plasma heating quickly leads to $v_S\ge
v_A$. Fast mode waves with the highest phase speed should dominate
the SA processes. We assume that they dominate the energy dissipation as the turbulent flow moves away from the SF and $v_F$ evolves from $v_A$ to $u$ at the same time. For a fully ionized
hydrogen plasma with isotropic particle distributions, which are
reasonable in the absence of strong large scale magnetic fields,
the TTD rate is given by \citep{s62, q98, pyl06}
\begin{equation}
\Lambda_T(\theta, k) = {(2\pi k_{\rm B})^{1/2}k\sin^2\theta\over
2(m_e+m_p)\cos\theta}\left[\left(T_em_e\right)^{1/2}\exp
\left(-{m_e\omega^2\over 2 k_{\rm B} T_ek_{||}^2}\right)+
(T_pm_p)^{1/2}\exp\left(-{m_p\omega^2\over 2k_{\rm B}
T_pk_{||}^2}\right)\right]\,, \label{d1}
\end{equation}
where $T_e$, $T_p$, $m_e$, $m_p$, $\theta$, $\omega$, and
$k_{||}=k\cos\theta$ are the electron and proton temperatures,
masses,  angle between the wave propagation direction and mean magnetic
field, wave frequency, and parallel component of the wave vector,
respectively. The first and second terms in the brackets on the
right hand side correspond to damping by electrons and protons,
respectively. %When $v_A\le 1.9\, v_S$, the electron and proton
%heating rates by fast mode waves are comparable
For weakly magnetized plasma with $v_A<v_S$, proton heating always
dominates the TTD for $\omega^2/k_{||}^2\sim v_S^2\sim 2k_{\rm B}
T_p/m_p$ (Quataert 1998). If $v_A$ does not change dramatically in
the downstream, the continuous heating of background particles
through the TTD processes makes $T_p\rightarrow (m_p/m_e) T_e$ since
the heating rates are proportional to $(mT)^{1/2}$, where $m$ and
$T$ represent the mass and temperature of the particles,
respectively. Parallel propagating waves with $\sin
\theta=0$ are not subject to the TTD. Obliquely
propagating waves are damped efficiently by the background
particles. Although the damping rates for waves propagating nearly
perpendicular to the magnetic field with $\cos\theta \simeq 0$ are also
low, these waves are subject to damping by magnetic field wandering
\citep{pyl06}. The turbulence power spectrum cuts off sharply when
the damping rate becomes comparable to the turbulence cascade rate
$\Gamma = \tau_{edd}^{-2}/(\tau^{-1}_F +\tau^{-1}_{edd})\simeq
\tau_{edd}^{-2}\tau_F$ \citep{j08}.  One can define a critical
propagation angle $\theta_c(k)$, where $\Lambda_T(\theta_c, k) =
\Gamma(k)$. Then for %$k_{\rm B} T_e = C_{2e} m_p u^2$ and
$k_{\rm B} T_p=C_{2p} m_p u^2$ with $C_{2p}< 3/5$ for $v_F<u$,
equations (\ref{IK}) and (\ref{d1}) give
$$ {v_A^2 k_d^{1/2} \over 2^{1/2}\pi^{3/2}  C_{2p}^{1/2} u v_F k^{1/2}} %\nonumber \\
\simeq  {\sin^2\theta_c\over \cos\theta_c} \exp\left(-{v_F^2 \over
2C_{2p} u^2\cos^2\theta_c}\right) \,.$$ where the electron pressure and damping
have been ignored.

For $v_A\ll U$, we have $3v_F^2/5\simeq v_S^2 \simeq C_{2p} u^2 =
[3C_{2p}/16(C_{2p}+1)]U^2$. The acceleration and scattering times of
relativistic electrons by these nearly parallel propagating waves
are energy independent and given, respectively, by $\tau_{sc} =
{2\pi/c k_d}\,,$ and $\tau_{ac} = ({3c^2/v_F^2})\tau_{sc} \,$
\citep{l06, be87}, where $c$ is the speed of light. The electron escape time from the large scale eddies is given by $\tau_{esc} = L^2/4c^2\tau_{sc}$, and
the spectral index of the accelerated electrons in the steady state
is given by $ p = \left({9/4} +
{\tau_{ac}/\tau_{esc}}\right)^{1/2}-{1/2}
%= \left({9/4} + {12c^2 k_m^2/ v_F^2 k_d^2}\right)^{1/2}-{1/ 2}
=\left({9/ 4} + {12c^2v_A^8
/ C_1^6 u^6 v_F^4 }\right)^{1/2}-{1/ 2}\,. $ The maximum energy that
electrons can reach %though resonant interactions with these parallel propagating waves
is given by $\gamma_cm_ec^2=2\pi qB/k_d%= qBL v_A^4/C_1^3u^3v_F =qBL\{[(p+0.5)^2-2.25]/12\}^{1/2}(v_F/c)
=7.26[(p+0.5)^2-2.25]^{1/2}
[C_{2p}/(C_{2p}+1)]^{1/2}$ $(L/10^{18}{\rm cm})(B/10 \mu{\rm
G})U_0$ TeV, where $U_0=U/0.015c$, $q$ is the elementary charge units. The
ratio of the dissipated energy carried by non-thermal particles to
that of the thermal particles should be greater than $\eta =
\theta_c^2(k_d)/2 = e^{5/6}v_A^2/2\pi(2\pi C_{2p})^{1/2} u v_F$,
where $e=2.72$, since the isotropic turbulence with $k<k_d$ can also
accelerate particles with the Lorentz factor $\gamma\ge \gamma_c$.
%Considering the evolution of the supersonic flow from $C_{2p}\sim v_A^2/u^2\ll1$
%to $C_{2p}\simeq3/5$, we estimate an acceleration efficiency of $\sim
%(20/3)^{1/2}\eta$.
%Because lower energy electrons and protons resonate with
%right-hand and left-hand polarized waves, respectively, we expect
%that nonthermal electrons and protons carry comparable energy.
%It is interesting to note that the spectral index of the accelerated
%particles are intimately connected to the shock speed and the SA
%efficiency $p = [9/4 + 12(2\pi)^6C_{2p}^2 c^2\eta^4/e^{10/3} C_1^6
%u^2]^{1/2}-1/2 = [9/4 + (4\pi)^6 C_{2p}^2(C_{2p}+1)
%c^2\eta^4/e^{10/3} C_1^6 U^2]^{1/2}-1/2$ and $\eta =
%\{[(p+0.5)^2-2.25]e^{10/3}C_1^6U^2/(4\pi)^6C_{2p}^2(C_{2p}+1)c^2\}^{1/4}$.

Electrons with even higher energy interact with the nearly isotropic
fast mode waves at $k<k_d$. The corresponding acceleration and
escape times are proportional to $\gamma^{1/2}$ and $\gamma^{-1/2}$,
respectively, which leads to a steady-state electron distribution
$f\propto \exp-(\gamma/\gamma_c)^{1/2}$ for $\gamma\gg\gamma_c$
\citep{pp95, bld06}. To obtain the complete electron distribution
from the thermal energy to the high-energy cutoff self-consistently,
one needs to obtain the acceleration and scattering rates by these nearly parallel propagating (with $k>k_d$) and
nearly isotropic (with $k<k_d$) fast mode waves and solve the
particle kinetic equation numerically. These are beyond the scope of
this paper. Instead, the above discussions indicate that the
particle distribution may be approximated reasonably well by
$f\propto \gamma^{-p} \exp -(\gamma/\gamma_c)^{\beta}$ with $\beta =
1/2$. To have significant particle acceleration, the acceleration
time needs to be shorter than the turbulence decay time:
$\tau_{ac}<\tau_{edd}(k_m)= L/C_1u$, which also validates the usage
of the steady-state solution of the particle kinetic equation and
implies  %$v_A= \{[(p+0.5)^2-2.25]C_1^6u^6v_F^4/12c^2\}^{1/8}< (C_1^2u^2v_F^3/3c)^{1/4}$, i.e.,
$C_1<(20C_{2p})^{1/2}/3
[(p+0.5)^2-2.25]^{1/2}$.
%For $C_2=1/2$ and $p=2$, this leads to $C_1<1/3^{1/2}$. In general,
%one needs to consider the time dependent solution for the accelerated particles \citep{bld06}.

\section{Results}
\label{app}

\begin{figure}
\plotone{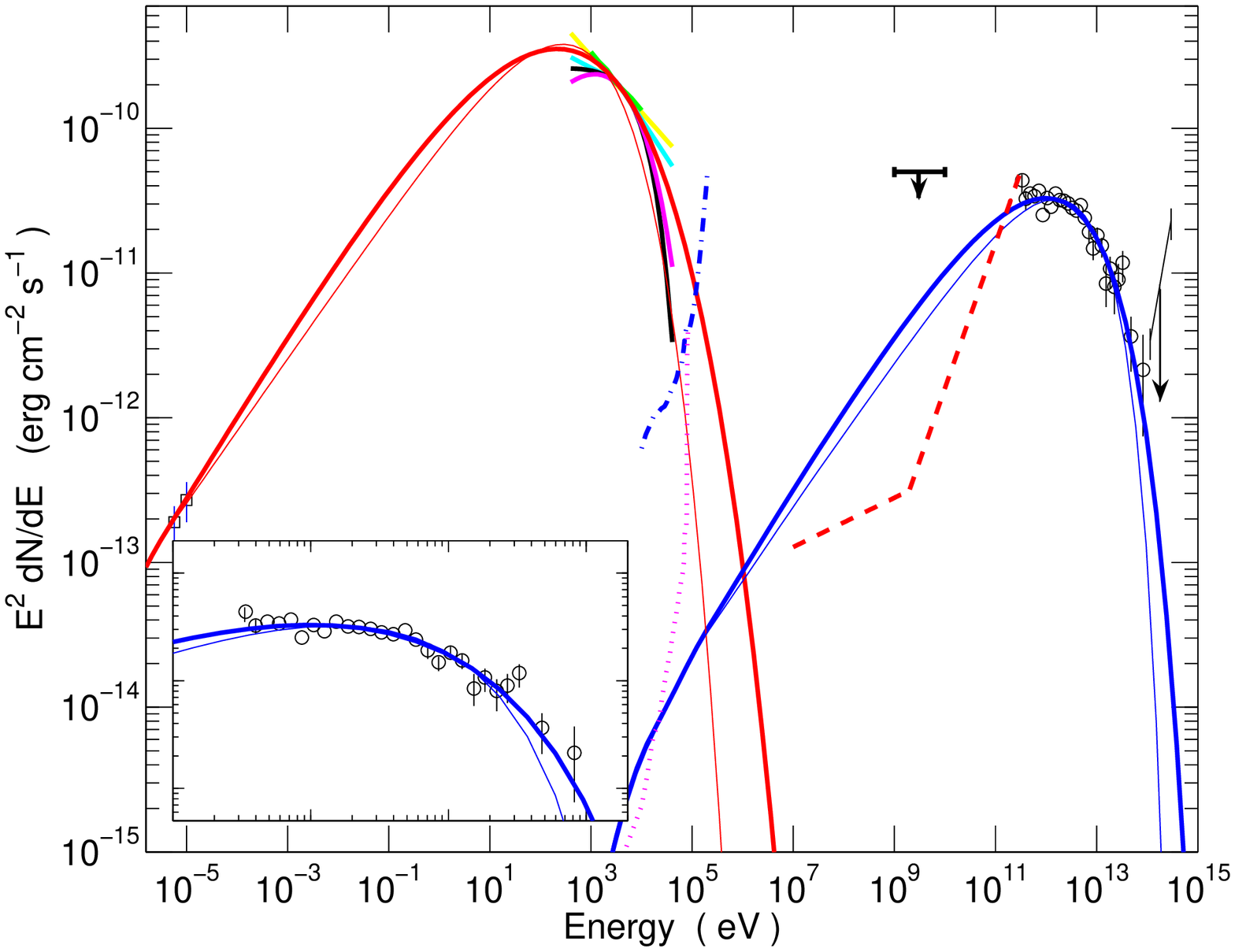} \caption{Model fit to the broadband spectrum of SNR
RX J1713-3946. The low and high energy spectral humps are produced through the synchrotron and IC scattering of the Galactic interstellar radiation, respectively. The X-ray data are for the four models discussed in
\citet{t08a} and rescaled to the {\it ASCA} flux level for the whole
remnant \citep{u03}. The thick solid line corresponds to the
fiducial model with $\beta=0.5$, $p=1.85$, $\gamma_c = 7.2\times
10^6$, and $B=12.0\mu $G. The thin solid line with $p=2.0$, $B=12.0\mu $G, and $\gamma_c = 4.4\times
10^7$ is the best for an electron distribution with an exponential
cutoff, i.e., $\beta=1.0$. The fiducial model gives a better fit to
the X-ray and TeV spectra. Insert is an enlargement of the TeV
spectrum to demonstrate the improvement of the fitting with the
fiducial model. The dashed line gives the {\it GLAST} integral
sensitivity (1 yr $5\sigma$: McEnery et al. 2004). The dot-dashed
and dotted lines correspond to the {\it HXMT} ($10^5$s $3\sigma$: Li
et al. 2006) and {\it NuSTAR} ($10^6$s $3\sigma$:
www.astro.caltech.edu/$\sim$avishay/zwicky1/ppts/Harrison.ppt)
sensitivities, respectively. \label{fit}}
\end{figure}

\begin{figure}
\plotone{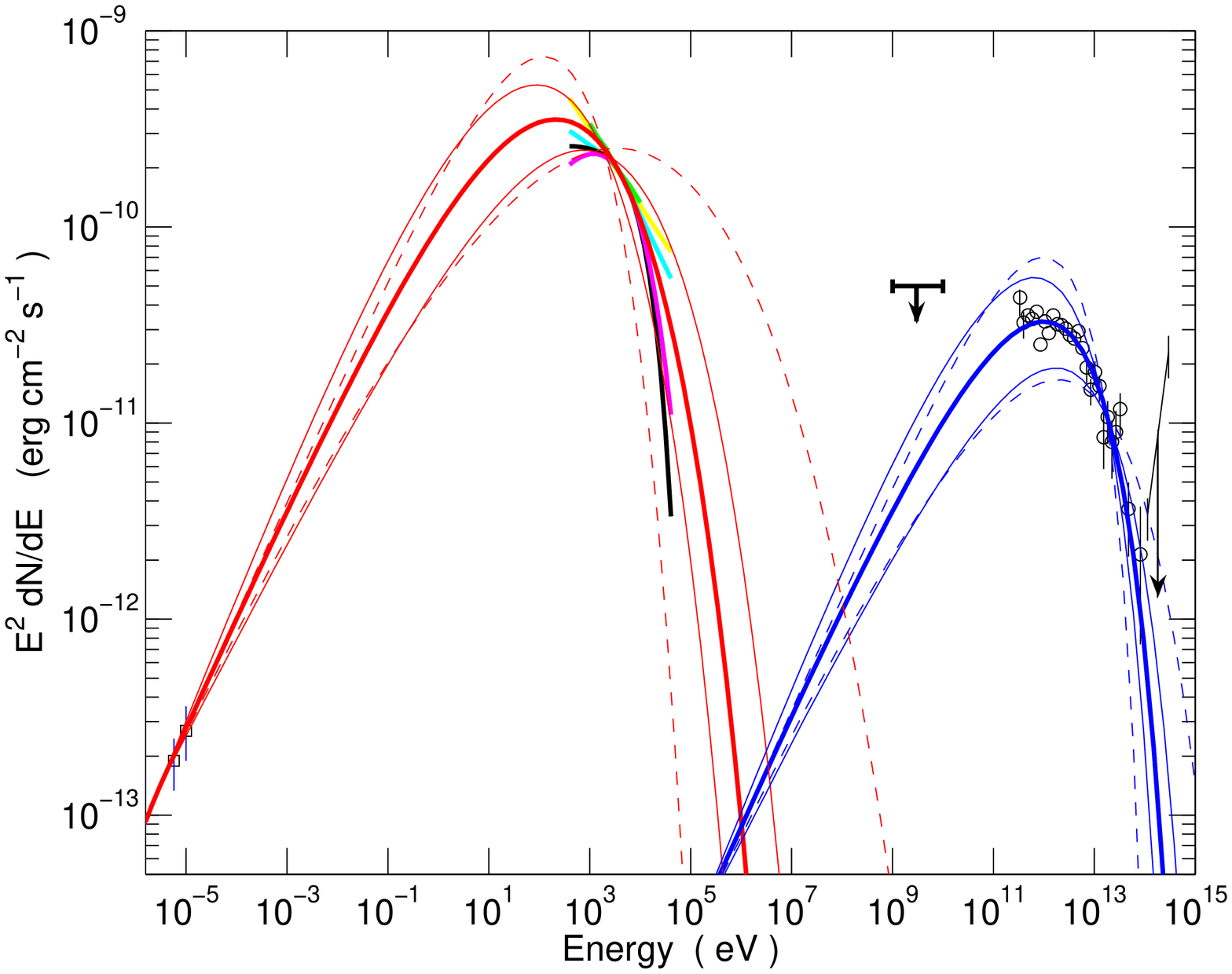} \caption{Dependence of the emission spectrum on
$\beta$ and $p$. $B=12$ G. $\gamma_c$ and $E_e$ are adjusted to fit
the radio to X-ray fluxes. The two dashed lines are for $p = 1.85$
and $\beta= 0.25$ and $1.0$. The thin solid lines are for
$\beta=0.5$ and $p = 1.7$ and $2.0$. \label{dep}}
\end{figure}

Several STSNRs have been observed extensively in the radio, X-ray,
and TeV bands. X-ray observations with {\it Chandra}, {\it
XMM-Newton}, and {\it Suzaku}, and TeV observations with HESS have
made several surprising discoveries that challenge the classical
diffusive shock particle acceleration model. The SNR RX J1713.7-3946
is about $t =1600$ years old \citep{w97} with a radius of $R\simeq
10$ pc and a distance of $D \simeq 1$ kpc.  By fitting its broadband
spectrum with the above electron distribution and background photon
field given by \citet{p06} through the synchrotron and IC processes, we find that $p = 1.85$, $B = 12.0\;
\mu$G, $ \gamma_c m_e c^2 = 3.68$ TeV, and the total energy of
electrons with $\gamma>1800$ \citep{l06, f08}, $E_e=3.92 \times 10^{47}$
erg (Fig. \ref{fit}). %These values are in agreement with previous studies \citep{p06, a06}.
The thin line shows the best fit with $\beta=1$, whose TeV spectrum is
not broad enough to explain the HESS observations. In the hard X-ray
band, our model predicted emission spectrum is significantly harder
and brighter, which gives better fit to the {\it Suzaku}
observations and can be tested with future {\it HXMT} and {\it
NuSTAR} observations. Our model also predicts a higher MeV flux
testable with future {\it GLAST} observations. Figure \ref{dep}
shows the dependence of the spectrum on $p$ and $\beta$.

The distribution of relativistic electrons obtained above implies
$L= 2.34 \times 10^{17} [C_{2p}/(C_{2p}+1)]^{-1/2}U_0^{-1}$
cm, and the electron density $n_e %= 12^{1/4}c^{1/2} B^2/4\pi (1+2Y)m_p[(p+0.5)^2-2.25]^{1/4}(C_1u)^{3/2}v_F$ $
= 2.00\times
10^{-3} (C_{2p}+1)^{5/4}C_1^{-3/2}C_{2p}^{-1/2} U_0^{-5/2}$
cm$^{-3}$, where $Y\simeq 0.1$ is the Helium abundance. For the
steady-state solution to be applicable, $C_1<0.824
C^{1/2}_{2p}<0.64$. For $C_{2p}=C_1=0.5$, we have $L =
4.04\times10^{17}U_0^{-1}$ cm, $n_e = 1.33\times
10^{-2}U_0^{-5/2}$ cm$^{-3}$, $\tau_{ac} = %6\pi c/v_F^2k_d = 3^{1/2}[(p+0.5)^2-2.25]^{1/2}L/2 v_F =
138 U_0^{-2}$ yrs,
$\tau_{esc} %= \tau_{ac}/[(p+0.5)^2-2.25]
= 42.3 U_0^{-2}$
yrs, and $\eta = 0.384\%U_0^{1/2}$. The energy
carried by the magnetic field $E_B = B^2R^2L/2 =2.77\times 10^{46}
U_0^{-1}$ erg. The short coherent length of the magnetic
field $2\pi/k_d= \gamma_cm_ec^2/qB=1.02\times 10^{15}$ cm is
consistent with the low level of polarization from most part of the
radio image \citep{l04}.
If $k_d$ doesn't change significantly in the downstream, the distance that
relativistic electrons diffuse through the $t = 1600$ yr lifetime of
the remnant is about $(t/\tau_{esc})^{1/2} L= 0.805$ pc. Most of them are
therefore trapped near the SF. The total kinetic energy carried by
the supersonic dissipative layer in the lab frame is $K \simeq 2\pi
R^2 L (1+2Y) m_p n_e (3U/4)^2 = 7.34\times 10^{48}U_0^{-3/2}$
erg. The corresponding gas mass $M = 2K/(3U/4)^2 = 1.29\times
10^{32}U_0^{-7/2}$ g,  thermal energy $E_t \simeq 6\pi R^2 L
m_p n_e(1+2Y)v_S^2 %= K/2
= 3.67\times 10^{48} U_0^{-3/2}$
erg, gas temperature $k_{\rm B}T_p = 2m_p(1+2Y) E_t/3M(1-Y) = 26.4
U_0^{2}$ keV, where we have assumed that the temperatures of
protons and Helium ions are the same and $T_e\ll T_p$.

The acceleration efficiency of $\sim\eta$ is not sensitive to $p$ and
$U$ and therefore doesn't change significantly through the evolution
of the SNR, implying that a thermal energy of $E_e/\eta \sim
1.02\times 10^{50}U_0^{-1/2}$ erg, which is $27.8$ times
higher than $E_t$, be produced in accompany with the electron
acceleration. The mass of the shocked thermal plasma is about
$E_eM/\eta E_t\simeq 3.58\times 10^{33}U_0^{-5/2}$ g$=
1.79 U_0^{-5/2}M_\odot$. These suggest that most of the
relativistic electrons were accelerated more than $\tau_{ac}\sim100$
years before outside the current dissipative layer. If the shocked
plasma is uniformly distributed within a shell of 4.5 pc (Aharonian
et al. 2006), the electron density is $1.74\times 10^{-2}U_0^{-5/2}$
cm$^{-3}$, which is comparable to that inferred from X-ray
observations (Cassam-Chena\"{i} et al. 2004).
%For a 1D shock wave though a uniform medium, this implies that the
%remnant lifetime should be 43 times higher than the lifetime time of
%the dissipative shell $\tau_{edd} \simeq 197 U_0^{-2}$ years.
%This suggests $U>0.015c$, which appears to contradict with

However, X-ray observations of the north-west outer edge of the
remnant suggests $U<0.015c$ (Uchiyama et al. 2007). So the model
predicted thermal X-ray emission likely exceeds the observed value.
%For the realistic 3D supernova shock wave, the acceleration is more
%efficient in earlier phase where both the shock speed and the gas
%density is high. Therefore a shock speed of $0.015c$ may still be
%acceptable. On the other hand,
The north-west edge is interacting with dense molecular clouds. It
is possible that the shock speed in a low density medium is still
greater than the observed upper limit so that $U\ge 0.015 c$. On the
other hand, given the uncertainties in the collisionless electron
heating processes, the electron temperature may be well below $1.0$
keV so that no significant thermal X-ray emission is expected in the
{\it XMM-Newton} and {\it Chandra} X-ray bands. Coulomb collisions
will make $k_{\rm B}T_e \simeq 0.198 (n_e/0.01{\rm cm}^{-3})^{2/5}
(t/1600{\rm yr})^{2/5} U_0^{4/5}$ keV \citep{h00}. Line
emission dominates the cooling with a luminosity of $\sim 10^{33}$
erg s$^{-1}$ and may be observed in the UV to soft X-ray band.
Detailed modeling of the evolution of the SNR is needed to reach
more quantitative results \citep{d00}.

X-ray observations discover thin bright features varying on a
timescale of about a year. The width of these features is about
$w\sim 4\farcs0\simeq 6.0\times 10^{16}$ cm, corresponding to a
speed $\ge 0.063 c$. Since the shock speed is much less than this
value, the variability has been attributed to cooling in very strong
magnetic fields \citep{u07}. According to our model, the decay may
be due to the diffusive escape of relativistic electrons from bright
regions.  For electrons with a scattering mean free path of
$2\pi/k_d$, the corresponding diffusive escape time from a region of
size $w$ is about 0.93 years. The X-rays are produced by electrons
with energies greater than $\gamma_c m_e c^2$, whose scattering mean
free path by fast mode waves should be longer than $2\pi/k_d$ implying a diffusion timescale of a few months. The
formation of these bright features is then related to the
inhomogeneity in the pre-shock medium, which is supported by the
relatively higher level of radio polarization nearby \citep{l04}.
The fact that the brightening features are not located at the outer
edge of the remnant can be attributed to the projection effect.

We also apply the model to SNR RX J0852.0-4622. For $d=1$ kpc,
$R=17$ pc, we have $B=9.4\; \mu$G, $p=2$, $\gamma_c m_e c^2 = 2.94$
TeV ($\gamma_c=5.76\times 10^6$), and the electron energy $E_e=1.01 \times 10^{48}$ erg. These imply $L =
3.73\times10^{17}U_0^{-1}$ cm, $n_e = 7.76\times
10^{-3}U_0^{-5/2}$ cm$^{-3}$, $\tau_{ac} = 141
U_0^{-2}$ years, $\tau_{esc} = 35.3 U_0^{-2}$ years,
$\eta = 0.404\%U_0^{1/2}$, $K = 1.14\times
10^{49}U_0^{-3/2}$ erg, $M\simeq 2.01\times 10^{32}
U_0^{-7/2}$ g, $E_t = 5.71\times 10^{48} U_0^{-3/2}$
erg, $k_{\rm B}T_p = 26.4U_0^2$ keV, and $E_B = B^2R^2L/2
=4.53\times 10^{46} U_0^{-1}$ erg. These two remnants are
very similar.

\section{Conclusions}
\label{con}

The steady-state distribution of relativistic electrons accelerated
by fast mode waves in a downstream turbulent dissipative
layer with a thickness $L$ of weakly magnetized collisionless astrophysical shocks can be
approximated as $$f\propto
\gamma^{-p}\exp-(\gamma/\gamma_c)^{1/2}\,,$$ where $p =
(9/4+2^{25}3^3c^2v_A^8/5^2U^{10})^{1/2}-1/2$,
$\gamma_cm_ec^2=qBL5^{1/2}[(p + 0.5)-2.25]^{1/2}U/24c$ for $
C_1=C_{2p}=0.5$, and the ratio of the dissipated energies going into
nonthermal and thermal particles is $\sim\eta =
(e^{5/3}96/5\pi^{3})^{1/2}v_A^2/U^2.$ For $p>2.09$, the
turbulence decay time is shorter than the particle acceleration time
and one needs to consider the time dependent solution. %The SA is
%negligible in the subsonic phase, when the free energy carried by
%the turbulence is converted into heat gradually.
The accelerated
electrons in the STSNRs give excellent fits to the broadband spectra of
SNRs RX J1713.7-3946 and J0852.0-4622. The model attributes the
recently observed variable X-ray features to inhomogeneity in the
upstream and the spatial diffusion of relativistic electrons near
the cutoff energy and predicts harder hard X-ray spectrum, higher
hard X-ray and $\gamma$-ray fluxes than leptonic models with a
sharper cutoff in the electron distribution. Future high sensitivity
hard X-ray and $\gamma$-ray observations will test the
model.
%Although the acceleration efficiency may be increased by the first-order Fermi process, our results suggest that the SA alone can explain observations of the STSNRs.
%The nearly parallel propagating fast mode waves are right-hand polarized and cannot accelerate protons and ions from the background plasma \citep{pl04}. However, given the efficient heating of protons through the TTD processes, some of the background ions may be accelerated to high energies through the TTD, providing the injection for further acceleration through the first order Fermi process. This ion acceleration may affect the overall energy budget near the SF significantly \citep{h00}.
The major uncertainty of the model is related to the turbulence generation mechanism near the collisionless SF. In this paper, we assume that the turbulence is isotropic and has a characteristic eddy size of $L$. Little is known about the evolution of anisotropic turbulence. The latter is valid as far as the free energy dissipation through shocks proceeds over the scale $L$, instead of, instantaneously at a narrow region ($\ll L$) near the SF as suggested in the diffusive shock models.
%It is obvious that identification of proper turbulence generation mechanisms plays essential roles in making the theory more self-consistent and robust.

\acknowledgments This work was supported in part under the auspices
of the US Department of Energy by its contract W-7405-ENG-36 to Los
Alamos National Laboratory and by the NSF of China (grants 10325313,
10733010, 10521001, and 10778726), CAS key project (grant KJCX2-YW-T03), the
Postdoctoral Foundation of China (grant 20070410636). We thank Dr.
Lu, F. J. for discussions on the instrument sensitivities.

\clearpage


\end{document}